# Magnetic liquid deformable mirrors for astronomical applications: Active correction of optical aberrations from lower-grade optics and support system

E. F. Borra[1]

Département de physique, de génie physique et d'optique,

Université Laval, Québec, Québec, CANADA, G1V 0A6

borra@phy.ulaval.ca

## ABSTRACT

Deformable mirrors are increasingly used in astronomy. However, they still are limited in stroke for active correction of high amplitude optical aberrations. Magnetic Liquid deformable mirrors (MLDMs) are a new technology that has advantages of high-amplitude deformations and low costs. In this paper we demonstrate extremely high strokes and inter-actuator strokes achievable by MLDMs which can be used in astronomical instrumentation. In particular, we consider the use of such a mirror to suggest an interesting application for the next generation of large telescopes. We present a prototype 91-actuator deformable mirror made of a magnetic liquid (ferrofluid). This mirror uses a technique that linearizes the response of such mirrors by superimposing a large and uniform magnetic field to the magnetic field produced by an array of small coils. We discuss experimental results that illustrate the performance of MLDMs. A most interesting application of MLDMs comes from the fact they could be used to correct the aberrations of large and lower optical quality primary mirrors held by simple support systems. We estimate basic parameters of the needed MLDMs, obtaining reasonable values.

*Subject headings:* Instrumentation: adaptive optics

## 1.   Introduction

During the past 25 years, deformable mirrors have been increasingly used for turbulence compensation at astronomical observatories (Duffner 2009). Presently, these mirrors are mostly made of solid thin plates or membranes and their stroke performance and number of actuators are sufficient at correcting atmospheric turbulence (Strachan et al. 2010). However, their stroke performance is insufficient for active correction of large surface or structural defects of the main telescope optics and support. Although some existing deformable mirrors can generate amplitudes of near a hundred microns, they can only do it for low order modes and are limited in the number of actuators that they can be produced with (Rooms et al. 2010). A promising technology to build a magnetic liquid deformable mirror that has sufficient stroke for large optical defects and number of actuators has been suggested by Borra et al. (2004). These magnetic liquid deformable mirrors offer an interesting alternative to solid ones because they do not require specialized fabrication equipment and are thus easy to build.

Magnetic liquid deformable mirrors (MLDMs) use a liquid (ferrofluid) made from a suspension of magnetic nanoparticles. Ferrofluids are deformed in the presence of a magnetic field and their surface profile is shaped by the geometry of the magnetic field. Compared to solid deformable mirrors, the main advantages of MLDMs come from the very high stroke and interactuator stroke that they can achieve (see section 2 of the present article). Their main disadvantage comes from the fact that they are constrained to remain horizontal. This is not a major limitation in a laboratory or at the telescope since the mirror can be placed at a stationary focus (e.g. Nasmyth or Coudé).

Astronomers should be aware of the possible astronomical applications of MLDMs that come from their technological advantages. For example, sub-optical system testing has become an important field in astronomical instrumentations. New giant telescope projects can be compared to space projects as they require ground based test support equipment to fully characterize their optical sub-system functionalities and performances before costly commissioning on the telescope. We recently demonstrated in Thibault et al. (2010) that MLDMs offer a credible and economical alternative to Computer Generated Holograms (CGHs) for optical system testing (e.g. telescope simulators). Following Borra et al. (2004) all of the subsequent work on MLDMs has been published in optics related journals. There has been considerable progress since then and MLDMs have acquired considerable credibility, positioning them as credible competitors to conventional solid deformable mirrors (Brousseau et al. 2010).

The novel suggestion in this article makes use of the large stroke and inter-actuator stroke that MLDMs can achieve. Here, we suggest that they could be used as active optical correction devices, reducing the error budget requirements of the primary mirror of a telescope and its active support structure. Active correction of the optical aberrations induced by the primary mirror and support structure already exists on modern telescopes (Wilson et al. 1991; Salas et al. 1997), but the correction is done by an active support, not using a deformable mirror. One could also correct aberrations introduced by the telescope auxiliary optics. As discussed in section 3, this would make it possible to build less expensive telescopes having good optical qualities. This cost issue is important, because, although not a technical issue, it is in practice the major problem that limits the availability of telescopes to astronomers.

Because this is the first time that the use of MLDMs to correct very large defects has been suggested, we cannot delve into details since it would require too much effort and too long an article. We simply experimentally demonstrate the very large amplitudes required, demonstrate the correction of the defects of a large mirror during the polishing state, list some of the issues involved and discuss them briefly. The technique may not be able to give a diffraction limited instrument nor allow a large field of view. The technique possibly may also be more useful for specialized inexpensive telescopes than general purpose telescopes.

## 2. Magnetic liquid deformable mirrors and experimental results

### 2.1. Basics

Since Borra et al. (2004) progress on the basic technology of MLDMs has been reported in the scientific literature and we give here a brief review for convenience. Ferrofluids are liquids that contain a dispersed suspension of magnetic nanoparticles that have small diameters (about 10 nm). When subjected to an external magnetic field the magnetic particles align themselves along the magnetic field and the liquid responds by changing its shape in a way to minimize the energy of the system. Consequently, any desired shape can be produced at the surface of the liquid as long as one is using the appropriate magnetic field geometry. This can be done by means of current carrying networks configured as straight wires (Borra et al. 2008) or with small coils (Brousseau et al. 2010). During the last few years, we have concentrated our work on MLDMs that use arrays of small coils.

Ferrofluids have low reflectivities (about 4%) and therefore need a reflective coating for most astronomical applications. We presently are using coatings made of colloidal silver nanoparticles called MeLLFs (Metal Liquid-Like Films) (Borra et al. 2004). The reflectivities of MeLLFs have improved considerably since Borra et al. (2004) as can be seen in Faucher et al. (2008). Since MeLLFs are not compatible with commercial ferrofluids, we have developed custom ferrofluids having characteristics necessary for compatibility with MeLLFs (Dery et al. 2008). We also have shown that certain liquids can be coated using surface deposition techniques that could be adapted to ferrofluids (Borra et al. 2007). The pour point of most ferrofluids is in the $-95^\circ$C range, this would allow using MLDMs at low temperatures.

The maximum surface deformation that can be obtained by ferrofluids is limited by the Rosensweig instability (Rosensweig 1997). In the worse-case scenario, i.e. just before the onset of the instability, one could produce deformations of the order of a millimeter using a non-coated commercial ferrofluid. Note that the reflective coating increases the value at which this onset appears, allowing much larger strokes than this, as can be seen in Fig. 1. For MLDMs, the magnetic field is given by the sum of the magnetic field components from each individual coil. As the coils are made of several current loops, the magnetic field they produce can be easily computed. Numerical simulations and experimental data show that the response of a single coil is well approximated by a Gaussian profile and that the full width at half maximum (FWHM) of the deformation can be tuned by changing the distance between the ferrofluid surface and the top end of the coils.

Until recently, some limitations with MLDMs were present. Firstly, the vectorial behavior of the magnetic field prohibits the use of standard techniques to predict the surfaces produced by the MLDM. Secondly, since deformations are proportional to the square of the applied magnetic field, only positive deformations can be produced on the MLDM surface. Thirdly, to reduce the current requirements, ferrite cores were used inside the coils, rendering the predictability of the MLDM performance complex and leading to hysteresis effects. A method to control MLDMs that eliminates these drawbacks has recently been proposed by Iqbal & Amara (2007). The technique uses a constant magnetic field, superimposed to the magnetic field of the coil array, significantly larger than the magnetic field of a single coil. The magnetic field of the coils becomes a small local perturbation of the external coil magnetic field. This linearizes the response of the coils (surface deformation becomes linearly proportional to the coils magnetic field) and also removes the vectorial behavior. The external magnetic field also amplifies the maximum amplitude that the mirror can produce. This was experimentally demonstrated by Iqbal & Amara (2007) and later confirmed by Brousseau et al. (2010). This technique allows us to use regular singular value decomposition (SVD) algorithms. It also makes negative deformations possible; thereby doubling the available stroke of the mirror without increasing residual errors that would be introduced by the need of a bias deformation on the mirror. Finally, as the small magnetic field of each coil gets multiplied by the external magnetic field, this reduces the current requirements of the coils so that ferrite cores inside them are no longer necessary. This new development thus greatly simplifies the usage of MLDMs and facilitates miniaturization.

## 2.2. A 91-actuator MLDM prototype

The MLDM prototype we currently use consists of 91 2.8-mm diameter custom coils arranged in a 33-mm diameter hexagonal geometry (Brousseau et al. 2010). Each resin-coated coil consists of about 300 turns of AWG36 magnet wire (0.13 mm in diameter) wound on a 1-mm diameter brass core. The coils are supplied in current by a custom amplifying stage controlled by an analog output PCI card.

A Maxwell bobbin is used to produce the uniform and constant magnetic field surrounding the coils. A Maxwell bobbin is an offshoot of a Helmholtz bobbin made from 3 coils instead of 2, giving a magnetic field of higher uniformity inside the region near its center than its Helmholtz counterpart (Maxwell 1873). The magnetic field produced near the center of the Maxwell bobbin is about 40 gauss when it is supplied with the 0.8 A current that is used while working with the MLDM. That gives a ratio between the magnetic field from the Maxwell bobbin and a single coil from the array of about 40.

The coil array is placed within the Maxwell bobbin with the top portion of the actuators lying near the middle of the bobbin form, exactly where the magnetic field is uniform. A container filled with a 1-mm depth of ferrofluid sits on top of the coils and a circular 50-mm diameter optical-quality BK7 window covers the container. The BK7 window is used to protect the liquid surface from dust particles and air currents from the room air exchange system. The window does not introduce chromatic aberrations because the mirror is lighted by a 659.5-nm laser diode. The window also does not introduce significant optical aberrations to the optical setup when the liquid surface is at rest but slightly affects the recorded wavefront of the mirror when it is driven because of its thickness. Note also that in an optimized setup, the liquid surface could be protected using a very thin nitrocellulose membrane which doesn't introduces optical path variations, second surface reflections and chromatic aberrations. A Shack-Hartmann wavefront sensor having a 44 x 44 lenslet array is used for the wavefront measurements. As the liquid surface must remain horizontal, a fold mirror is used to image the liquid surface. A 1∕5X telescope is used to image the wavefront produced by the MLDM to the 5-mm pupil of the wavefront sensor. Note that vibrations were not found to pose a problem in our laboratory as the layer of ferrofluid used is thin. Furthermore, vibrations can be minimized by using a sufficiently viscous ferrofluid. Note that, in a telescope, the mirror could be protected by proper shielding. The glass window could also be replaced by a very thin nitrocellulose membrane that would have a negligible effect on the wavefront.

### 2.3. Experimental results

To construct the interaction matrix of the mirror, each coil is successively supplied a same current value and its influence function recorded using the wavefront sensor. A pupil size of 25 mm in diameter is used. The influence function slope measurements are used to construct the MLDM control matrix and the current signals required to produce a targeted surface are computed using singular value decomposition (SVD). Like with commercial deformable mirrors, once these measurements are performed, they should remain valid as long as one does not change the parameters of the mirror (e.g., the thickness of the ferrofluid layer). Fig. 2 shows a typical influence function of a single coil of the MLDM when driven at the maximum current available from the driving electronic. The figure shows that a 57 $\mu$m wavefront stroke response can be obtained from a single coil. This value could be raised by increasing the magnetic field of the external driving bobbin. Significantly higher amplitudes could also be reached with an optimized setup as discussed in the conclusion. As there is no strong physical constraint on the liquid surface, the inter-actuator stroke will always be of the same order of the stroke of a single coil. Fig. 3 shows the inter-actuator stroke produced when two neighboring coils are driven in opposite current directions. The figure shows an inter-actuator stroke of 57 $\mu$m that, again, could be increased by driving the external driving coil at a higher current. This inter-actuator stroke performance is much higher than the maximum inter-actuator stroke that can be produced by any known commercial DMs (Rooms et al. 2010).

Fig. 5a shows an astigmatism term having a 30 $\mu$m wavefront amplitude as an example of the kind of performance offered by MLDMs. Higher astigmatism amplitudes can be produced but we limited ourselves to 30 $\mu$m to maintain low residuals. Increasing the number of actuators would lower residuals at higher amplitudes. The residual wavefront error is shown at Fig. 5b and has a RMS value of 0.064 $\mu$m, corresponding to $\lambda/10$ at the system wavelength of 659.5 nm. Fig. 5c shows a wavefront residual prediction based on a simulation that uses the linear addition of purely Gaussian functions to describe the influence functions of the actuators. The amplitude and coupling constant (The coupling constant is the relative wavefront amplitude above an unpowered actuator when its nearest neighbor is active) of the Gaussian function used in the simulation were derived from Fig. 2 and set to 2.5 $\mu$m and 0.55. The RMS residual from the simulation is 0.059 $\mu$m, closely matching the experimental result both in amplitude and shape as seen by comparing Fig. 5b and Fig. 5c. This result is of great importance as it shows that we can accurately predict the performance of the mirror at producing specific wavefronts. It validates the use of our simulations in section 3.

We also produced astigmatism terms of 10 and 50 $\mu$m of wavefront amplitudes and obtained RMS residual error of 0.026 and 0.105$\mu$m. Both values are also in agreement with a linear model of the mirror. The RMS residual errors for other Zernike terms can be found in Brousseau et al. (2010), though they were produced at lower wavefront amplitudes. We can use a linear model of the mirror to predict MLDMs performance for higher number of coils. For example, Fig. 4 shows the predicted RMS residuals of a MLDM having 217 actuators and a coupling of 0.60 as function of Zernike index. Even if residuals are high for high order modes, they remain small for the dominating low order modes. Note that the steep increase of the residuals with increasing index is expected and due to the limited number of actuators and the large coupling constant of the MLDM.

The Earth's magnetic field vector can be considered constant over the small physical scale of the mirror. Large masses of ferrous metal can indeed modify the response of the mirror but these masses need to be in the immediate vicinity of the device as the influence diminishes as the third power of their distance. Moreover, ferrous masses further away will have a minor influence which can be corrected by the actuators.

## 3. Correction of the aberrations of large mirrors

Large optical telescopes are very expensive for two principal reasons: that the surfaces of their mirrors must be polished to high optical quality ($< \lambda/10$) and they must be held in very rigid, and/or adaptive, support systems (Wilson et al. 1991; Salas et al.

1997) so that they do not deform as the telescope moves around the sky. In principle there is an alternative far less expensive solution to the problem: One could correct the aberrations caused by lower-quality optics and an inferior support system using adaptive optics. However the problem now is that the aberrations will be impossible to correct with conventional deformable mirrors because, firstly one shall have to correct very high amplitude low-order deformations and, secondly, need a very large number of actuators. High inter-actuator strokes may also be needed depending on the aberration content. This will be needed, for example, in the case where the primary mirror is a segmented mirror made of an array of smaller mirrors (facets). There will then be discontinuities in the wavefront, due to misalignments among the facets. In the previous section we have shown that MLDMs are capable of producing extremely high inter-actuator strokes. Furthermore, because diameter of the reflective surface is not an issue with MLDMs, having a large number of actuators will be possible. We shall elaborate on this in the next two sub-sections, where we shall make use of two practical cases of existing poor-quality mirror and support system to estimate the parameters that are needed to correct their aberrations. These parameters will then be used to estimate the requirements of the active optics needed to correct the wavefront of a mirror made of poor-quality facets. We will then show that MDLMs are capable of the required performance.

### 3.1. Poor optical quality facets

One of the factors that makes an astronomical mirror expensive is that it has to be polished to better than $\lambda/10$. The early polishing (e.g. at an RMS of 10 microns) is relatively inexpensive but the cost increases considerably as one gets closer to $\lambda/10$ because one must carefully polish and frequently test the quality of the mirror surface.

Let us now consider the surface of a large mirror during the early polishing stages.

Our example is a large mirror undergoing polishing at the Steward Observatory mirror laboratory. They provided us the first 50 Zernike polynomials that characterize the surface of this mirror during the early polishing stage. Note that the optical quality within dimensions smaller than the polishing tool (e.g. a few tens of cm) is good and does not have to be corrected. Fig. 6a shows the reconstructed wavefront from these Zernike coefficients. The peak-to-valley amplitude is 55 $\mu$m and the RMS is 15.7 $\mu$m with, as expected, the defocus term clearly dominating. Correcting these relatively high amplitudes could easily be achieved using a MLDM. Figure 6b shows the predicted residual wavefront error after correction using a 217-actuator MLDM as in the preceeding section. After correction by the MLDM, the RMS is now 0.058 $\mu$m. Based on Marechal criterion ($\lambda/14$), the mirror is therefore diffraction limited after correction for wavelengths longer than about 800 nanometers. Although not diffraction limited for $\lambda < 800$ nanometers, the optical quality is still reasonable. Note that better correction could be achieved by using a larger number of actuators and/or tweaking the MLDM influence functions. For example, one can easily control the width of the influence function of MLDMs, by simply varying the thickness of ferrofluid or changing the distance between the liquid and the coils.

### 3.2.  The active support system

The active support system is the second factor that increases costs. To illustrate the advantage that our high-interactuator strokes give, we shall consider the extreme case of a Cherenkov telescope and show that MLDMs are capable of correcting the high amplitude aberrations introduced by their extremely poor support systems.

Cherenkov telescopes are used in gamma ray astronomy to observe the light generated by Cherenkov radiation. They are very large optical telescopes having diameters greater than 10 meters made of several hundreds of smaller mirrors (facets). Because the observed objects are extended and there is therefore no need for detailed images, the optical quality of the mirror is extremely poor by optical astronomy standards. Part of this poor optical quality comes from the inexpensive support system. There are no detailed quantitative analyses of the wavefronts produced by Cherenkov telescopes. However, we can obtain approximate useful estimates from published point spread function (PSF) profiles (Bernlohr 2003). Although the PSFs do not provide details about the relative contributions of the sources of aberrations, we can use these PSFs to make educated guesses about them.

We shall consider the published PSFs of the HESS telescope (Bernlohr 2003). At the zenith, the on-axis PSF is dominated by the aberrations of the poor-quality facets and has a width of 0.25 mrad ( 50 arcseconds). One cannot distinguish the contribution of the structure from the contribution of the facets themselves. However, the variation of the on-axis PSF with elevation is due to the deformation of the support dish. The deformation of the support becomes noticeable for zenith distances larger than 35 degrees. The PSF width never exceeds 0.6 mrad, even at zenith distances of 80 degrees.

Considering that the HESS PSFs are always smaller than 0.6 mrad, we shall assume in our simulations a telescope structure that generates aberrations that produce a PSF having a width of 1 mrad. On the basis of the published HESS PSFs (Bernlohr 2003), we shall make the reasonable assumption that the aberrations are dominated by low order aberrations in the hundred microns range. These amplitudes and low-order aberrations give PSFs comparable to those of the HESS. Note that we are not claiming that we can correct the PSFs of existing Cherenkov telescopes. In practice, this cannot be done because the facets have themselves extremely poor optical qualities and correcting the wavefronts of several hundreds facets would necessitate a prohibitively large total number of actuators. Instead, we simply shall use the structures of the Cherenkov to obtain estimates in the discussion that follows.

**3.3.   The case of a telescope that uses a primary mirror made of poor quality facets**

The estimates obtained in the previous two sub-sections shall now be used to consider a hypothetical telescope that uses a large primary mirror made of inexpensive smaller mirrors (facets) having optical qualities similar to those of the mirror during the polishing stage, discussed earlier in this section, that has defects with a peak-to-valley amplitude of 55 $\mu$m and an RMS of 15.7 $\mu$m. We shall also assume that the facets are held in a structure, similar to the structures of Cherenkov telescopes, having a performance similar to the one discussed earlier in section 3.2.

Table 1 gives the number of actuators required for various primary mirror sizes made of hexagonal arrays of 5-m and 8-m hexagonal facets. The results are based on the estimate that each facet requires at least 217 actuators to correct the aberrations induced by the primary mirror quality and support structure. The table also gives the full diameter of the needed MLDM. We assumed that each actuator has a diameter of 5 mm and that they are closely packed. This large diameter considers the fact that actuators larger than our 3-mm actuator diameter prototype would produce the required large deformations with both lower current consumption and lower heat generation. The table shows that the diameters of the MLDMs are reasonable and could readily be made using the same techniques used with our present mirrors. The numbers of actuators are comparable to existing projects of high-actuator-density deformable mirrors; however MLDMs are easier to fabricate with large number of actuators. This can be appreciated by considering that we made a 91-actuator mirror with the limted resources of a University laboratory. On the other hand the control complexity increases with the number of actuators. However, if one simply corrects the aberrations introduced by defects of the primary mirror and the support structure, the speed requirements are less stringent and the complexity problem is not as severe. Furthermore, there is ongoing work in adaptive optics to handle high actuator count DMs so that the problem will not be as severe in the future. In the discussion that follows, we estimate, on the basis of our own experience, that the cost of a MLDM should be less than \$80 per actuator. A $12,000$ actuators mirror would thus cost of the order of a million dollars, a negligible cost compared to the cost of a full 30-m telescope.

### 3.4. Comparison with Existing Active mirrors

The most recent telescopes (e.g. VLT and NTT) as well as the next generation of planned large telescopes (e.g. TMT and E-ELT) use active support systems (Noethe 2002). It is therefore legitimate to consider the advantages and disadvantages of our proposal compared to active support systems. The active optics of the VLT and NTT can clearly correct the existing defects of the primary mirrors. However, the defects, with the exception of easy to correct defocus, have low amplitudes. For example, the second mode $e_{2,1}$ of VLT has an amplitude of only 4 microns at a zenith angle of 45 degrees. Significantly higher spatial frequency modes would be difficult to correct and necessitate a high number of actuators. The actuators of these active systems are complex, must generate high forces and must be supported by a rigid underlying structure.

In comparison, a MLDM mirror could have a very large number of easy-to-make actuators (several tens of thousands or more) capable of strokes and inter-actuator strokes in the tens of microns so that it could compensate for very high spatial frequency and large amplitudes defects. This would allow one to have a primary mirror made of very thin facets held by simple support systems. This would also allow for a lighter and cheaper telescope frame.

Active optics have, over MLDMs, the advantage that they can compensate for discontinuities in the wavefronts of the facets, while MLDMs cannot. This is because the radius of the influence function gives a basic limit. On the basis of our experience with MLDMs, we know that a discontinuity is smoothed over a region having a diameter equal to twice the diameter of an actuator. Consequently, discontinuities in wavefronts can be corrected by mounting the facets so that there is an empty space between their edges. The empty space must have dimensions equal to the diameter of two actuators projected over the facets. The negative factor obviously is that it increases the diameter of the frame of the telescope since the facets cannot touch. We can estimate the effect by using table 1 and assuming hexagonal facets. If we consider the 35-m diameter mirror with 217 actuators per facet, we see that the diameter of the frame of the telescope would have to be increased by 20%, which is not a dramatic effect, considering the low cost of the frame (see next section ). Obviously, increasing the number of actuators would result in a smaller increase. The reduction would increase with the square root of the number of actuators. For example, using 868 actuators per facet would reduce the increase to 10%. One also could make a MLDM adaptive mirror with smaller actuators at the projected interfaces to minimize the problem. The problem will also be minimized with a small number of large facets. Based on the previous discussion, we can assume that the cost effect of increasing the primary mirror diameter will be small. This opinion shall have to be quantified by a detailed study.

## 4.    Discussion

The issues involved with the correction of the aberration of primary mirrors, including segmented mirrors, with adaptive optics has been previously considered (Yaitskova & Verinaud 2003) but only for the correction of the small amplitude defects that presently available deformable mirrors are capable. As can be seen in the above reference, a quantitative analysis of the issues with MLDMs would be too complex and is beyond the scope of this article. We will instead briefly discuss some of these issues and how one can deal with them.

The MLDM will need re-imaging optics which has inconveniences since they complicate the design, reduces the throughput, and add additional costs. This is however mitigated by the fact that one could use low-optical quality optics since their defects could be compensated by the MLDM. The exact tolerance allowable on the low-quality optics remains to be evaluated but will greatly depends on the exact configuration that one considers.

Adding a MLDM will reduce the throughput since it does not have 100% reflectivity. Presently MLDMs in our laboratory have peak reflectivity of the order of 80%. We are carrying work to improve this value. In our laboratory we find that MeLLFs added on water have reflectivity comparable to the reflectivity of solid silver. We can therefore expect that MeLLF-coated ferrofluids will eventually have comparable reflectivity. We are also working on MLDMs covered with a flexible membrane that is coated with an aluminum layer. Experimental results show that the influence functions are not noticeably affected by the membrane. Although the strokes are decreased by about a factor of three with respect to naked ferrofluids, membrane-coated ferrofluids are still capable of very large strokes.

Assuming that the MLDM perfectly corrects the wavefront originating from a star at the center of the field, it will not do it for the wavefronts originating from off-axis objects. This comes from pupil distortion and oblique reflections. Assuming that the primary mirror is perfectly imaged onto the MLDM for all rays originating at the center of the field of view, this would not be the case for rays originating off-axis because they strike the primary mirror and the MLDM at different angles. Consequently, the footprint of the pupil image on the MLDM changes with field angle. This is pupil image distortion that results in a misregistration of the MLDM actuators. The distortion increases with field angle. degrades the performance, and limits the telescope's useful field of view. These problems increase with telescope diameter and the stroke of the actuators of the deformable mirror and might be a limiting factor for very large telescopes. These problems can however be mitigated by the fact that MLDMs can be fabricated to very large diameters (meters). An estimate of the effect is given in the discussion that follows.

The misregistration between the MLDM actuators and the pupil image of the primary makes the design of the wavefront sensor critical as shown in Yaitskova & Verinaud (2003). As for the field of view, even if the deformable mirror corrects the wavefront at the center of the field, it is the ratio between the diameter of the primary mirror and the diameter of the deformable mirror that makes the deformable mirror overcorrect a target at the edge of the field. The problem is therefore minimized by using a large diameter deformable mirror. This is where the fact that MLDMs can be fabricated with large diameters plays an important role. We cannot use the MLDM in place of the secondary mirror because it is not tillable, so a careful examination of the design of a telescope using a MLDM is needed. A very large MLDM could for example be placed at the location of the Nasmyth platform. This will require large relay mirrors, increasing costs; but this cost increase will be mitigated by the fact that one could use low-optical quality mirrors since the wavefronts could be corrected by the MLDM. A quantitative estimate of these issues shall have to be made but is beyond the scope of this article. Nevertheless, we can make a crude approximation of the field of view offered by using a specific size of MLDM by considering the fact that the optical invariant will cause the angles in the MLDM space to be magnified by the magnifying ratio of the system. A piston error in the entrance pupil will thus be increased in MLDM space by the magnification factor and will be related to the cosine of the field angle. This will cause the rays at the edge of the field to remain uncorrected by $\delta(1 - 1/\cos m\varphi)$, where $\delta$ is the piston error for an on-axis ray, $m$ is the magnification factor between the equivalent diameter of the primary mirror and the deformable mirror, and $\varphi$ is the field angle (Bely 2003). If we take for example, a 30 m telescope, a 2.5 m diameter MLDM, and 200 waves of error in the entrance pupil, the previous equation shows that the field of view has an almost 20 arcminutes diameter for $1/10_{th}$ of a wavelength residual in the exit pupil plane.

Fig 5 shows that MLDMs are capable of producing deformations having high amplitudes (30 $\mu$m) and low residuals. Note however that we limited the amplitude to 30 $\mu$m because we wanted to obtain low residuals. Greater amplitudes can be achieved using MLDMs, as can be deduced by looking at Fig 2 and 3. The 30 $\mu$m limit was not imposed by the ferrofluid itself, but by the limits of our wavefront sensor as well as the high residuals that a wavefront generated with 91 actuators would have had at amplitudes larger than 30 $\mu$m. Thermal dissipation issues, mostly due to the external driving bobbin, were noted but only when pushing the MLDM to near its limit of operation (external bobbin running at 1.2 A), but this could be minimized by optimizing the design of the actuators and the external driving bobbin. External cooling would also help. Fig. 1 clearly shows that extremely large deformations (a few millimeters) are possible. The actual practical limit that comes from the heat generated by the currents needed to produce extremely large magnetic fields could also be minimized by ultra-low resistivity materials.

Building the MLDM used in our experiments was inexpensive. The whole MLDM, including the driving electronics, has an evaluated cost per actuator of about 180 US$. Half of this amount is due to having to buy low-demand off-the-shelf PCI cards from the manufacturers. The cost per actuator could be reduced by using custom-made electronics. Assembling the mirror was also relatively easy. This can be appreciated by noting that we made a deformable mirror with a rather large number of 91 actuators using the limited resources available in a typical University laboratory.

The cost issue is an important one but is very difficult to accurately quantify at this stage of the suggestion since we do not have a detailed design. We can only make assumptions. The cost will be somewhere between the cost of a currently planned large telescope and the cost of a Cherenkov telescope. At the lower limit, the cost of the mirror and telescope frame could be comparable to the cost of a Cherenkov telescope. The 17-m diameter MAGIC Cherenkov telescope has a total cost of 4.5 millions dollar (Lorenz 2004) including instrumentation. Recently Lorenz et al. (2010) have proposed a 23-m diameter Cherenkov telescope, made of 220 facets of 2 square meters area each, giving a point spread function with a diameter of 6 arcseconds for an estimated cost(including instrumentation) between 6 and 8 million Euros (8 and 11 million US dollars). They estimate that it would take 2 years to carry out required R&D and a detailed design with an additional 2 years for construction. The cost will however be higher for a telescope useful for astronomical observations since the structure may have to be more rigid. To that cost one must also add the cost of the MLDM (see section 3.3 for a cost estimate) relay optics and the cost of the enclosure. We shall not attempt to accurately quantify the cost related to these issues. It is very complicated since it depends on many parameters (e.g. number of segments, f-ratio, etc.). For discussion purposes, we can make the reasonable assumption that the cost of the MAGIC telescope must be multiplied by a factor of 5, giving us a 25 million dollar cost. To this one must add the cost of the housing enclosure. Let us now consider a simple enclosure; for example made of a square tower with a sliding roof. Assuming another 25 million dollars cost for this simple housing structure, we then reach a total cost of 50 million dollars. This cost estimate is obviously extremely approximate. We can then compare this estimate to the cost of a conventional 17-m diameter optical telescope by using the cost versus aperture relation in Fig 1 of (van Belle et al. 2004). The predicted cost would be of the order of 400 million dollars. As they state, the cost data points in their figure includes telescope mirror, structure, enclosure, and other essential site work, and is exclusive of instrumentation and operations cost.

One may however object that with the telescopes currently operating the primary mirror cost is about 25% of the total construction cost and about 10% of the total project cost (including 20-year operation) so that, at first sight, reducing the cost of the primary mirror seems to have a minor impact. However, the adaptive mirror correction would not only allow one to have a less expensive primary mirror but also give further cost savings because it would allow to use simpler and less expensive telescope frame and support system for the primary mirror. Furthermore the 25% relative cost is also due to the fact that the present large telescopes carry several instruments (e.g. a wide field of view camera, an adaptive optics system, spectrographs). These instruments are expensive and require high maintenance costs (e.g. staff) . The fact that in our case, the telescope is inexpensive would allow us to have specialized telescopes that carry a single instrument and will therefore be less expensive to operate.

## 5. Conclusion

The novel suggestion, presented in the previous section, that MLDMs can be used to correct the aberrations of large telescopes made of inexpensive facets held in inexpensive structures is particularly interesting because it can reduce the costs of the next generation of telescopes. The 30-m diameter telescopes presently under consideration will be so expensive that they will require international collaborations so that only a couple of them will be built and telescope time will obviously be limited and difficult to obtain. Less expensive telescopes could allow us to build several specialized telescopes, each having a single dedicated instrument (e.g. a high-resolution spectrograph), thereby giving us more total telescope time. It may also make it possible to build telescopes having diameters significantly larger than 30 meters. On the other hand, this type of telescope may not be as versatile as classical telescopes. For example, it may give point spread functions unsuitable for searches for extrasolar planets. Perfect correction of the on-axis wavefront may also be insufficient for imaging at the diffraction limit with adaptive optics.

Many issues have not been sufficiently addressed in this article since we only list them with a brief discussion. This will require considerable work. In particular, the cost issue shall have to be more accurately quantified after a more detailed design of such a system is performed.

The advantages offered by MLDMs (e.g. high-amplitude deformations, ease of fabrication and low costs) can have other astronomical applications. For example because MLDMs can produce large amplitudes and have a predictable performance, they are an interesting and innovative approach in sub-optical system testing (Thibault et al. 2010). Note that we do not discuss applications of MLDMs to correct atmospheric seeing. In principle, this could be done; however, we have not yet demonstrated a closed loop operation at sufficiently high frequencies.

This novel MLDM technology is still in its infancy and we can expect that it will improve considerably over the next few years. The reliability of this prediction can be appreciated by considering the progress made since Borra et al. (2004). The linearization technique discussed in section 2 gives a striking example of the progress made. Tiltable MLDMs are among the possible breakthroughs to expect because ferrofluids stick to magnets. We have applied a MeLLF-coated ferrofluid to a permanent magnet that we have then held in our hands and tilted in all directions (including the upside-down position). The coated ferrofluid stayed stuck to the magnet, showing a smooth (as evaluated by eye-inspection only) shiny surface. We did not evaluate the optical quality of this mirror with optical tests. We expect that it will take considerable work to develop a tiltable MLDM technology but we can speculate that eventually it might allow us to make adaptive secondary mirrors for astronomical telescopes. The millimeter size deformations seen in Fig. 1 cannot presently be attained with our mirror since the current needed would generate excessive heat. We can however speculate that the problem would be resolved by optimizing the design of the actuators and/or using ultra-low resistivity materials like Amperium™ wire recently introduced by the company American Superconductor®. It has the ability to conduct more than 100 times the electrical current of copper wire of the same dimensions. Using it would allow us to have 100 times larger strokes for the same heat generation. Furthermore, superconductors are currently the subject of major research efforts and we can expect that they will eventually be practical. Superconductors at the temperature of liquid nitrogen have already been demonstrated. This is an important breakthrough for technological applications of superconductivity because liquid nitrogen is an inexpensive and easy to handle coolant.

In parallel to the work presented within this article, effort is also put on the miniaturization of MLDMs in collaboration with micromachining experts from another institution.

This research was supported by the Natural Sciences and Engineering Research Council of Canada, the Canadian Institute for Photonic Innovations and NanoQuébec.

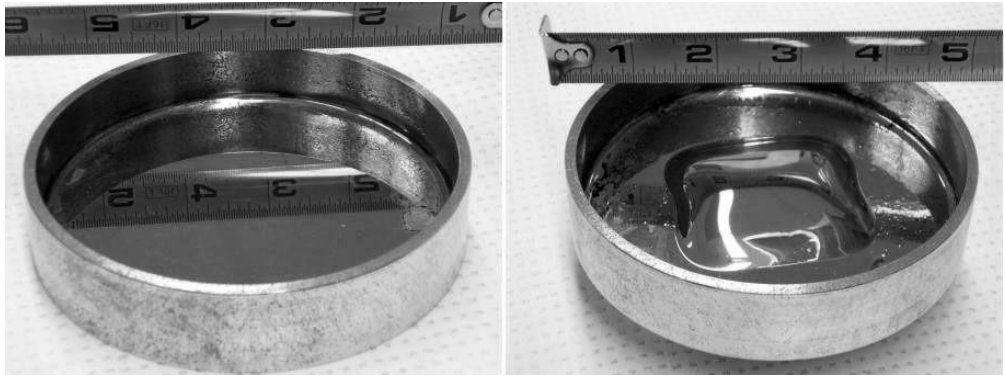

Fig. 1.— Photograph of a layer of ferrofluid submitted to the magnetic field of a small permanent magnet located under the container. A deformation of a few millimeter amplitude, visible to the naked eye, can be seen. The ferrofluid is coated with colloidal silver nanoparticules to render the surface reflective.

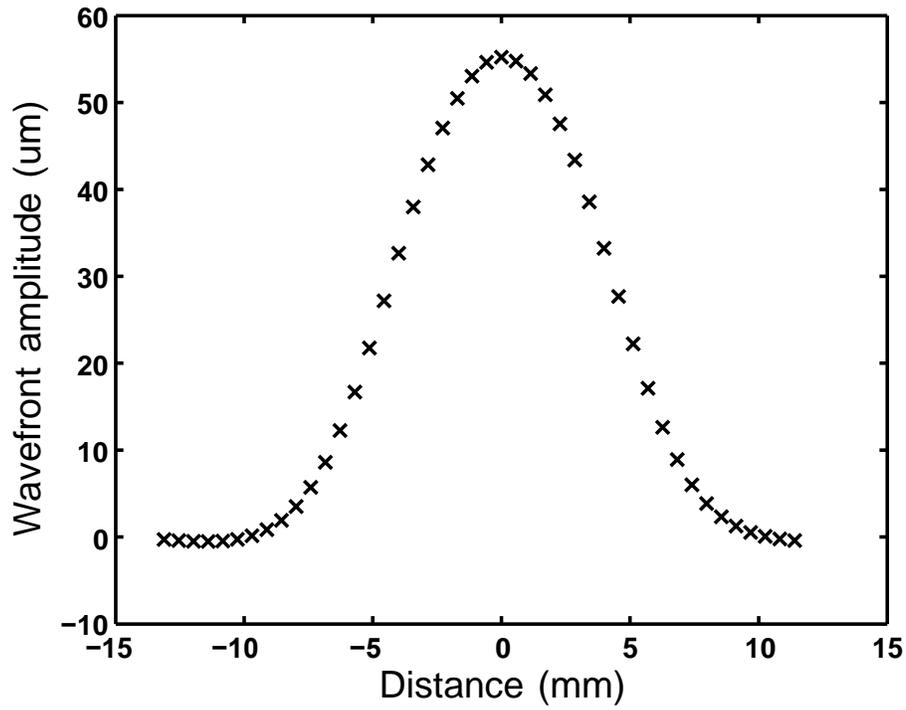

Fig. 2.— Influence function of a single actuator of a 91-actuator MLDM operating at maximum current available from the control electronic. The maximum wavefront amplitude could be further increased by using a higher current in the external coil that is used to linearize the response of the actuators.

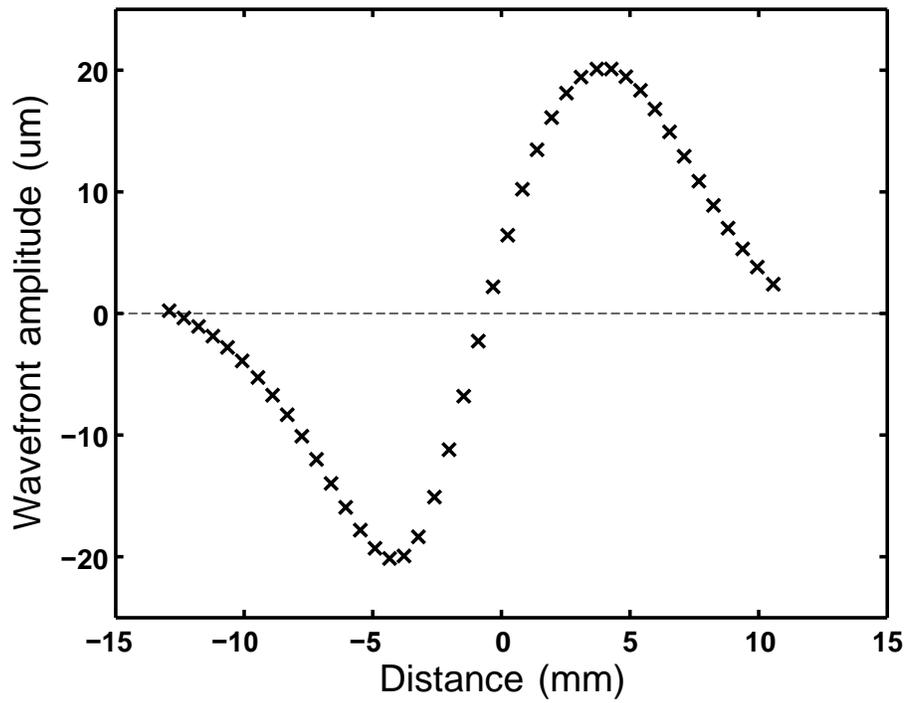

Fig. 3.— The inter-actuator stroke between two neighboring actuators of the MLDM. One actuator is driven with a negative current while its nearest neighbor is driven with a positive current.

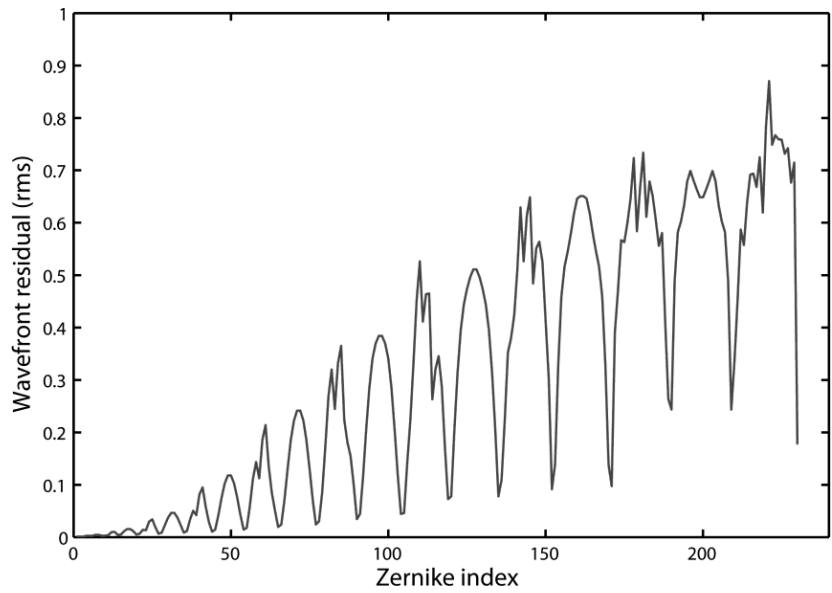

Fig. 4.— Predicted normalized RMS residuals of a MLDM having 217 actuators and a coupling of 0.60 as function of Zernike index. Note that the steep increase of the residuals with increasing index is expected and mostly due to the limited number of actuators and the large coupling constant of the MLDM. The vertical scale is normalized to a full correction of 1.0.

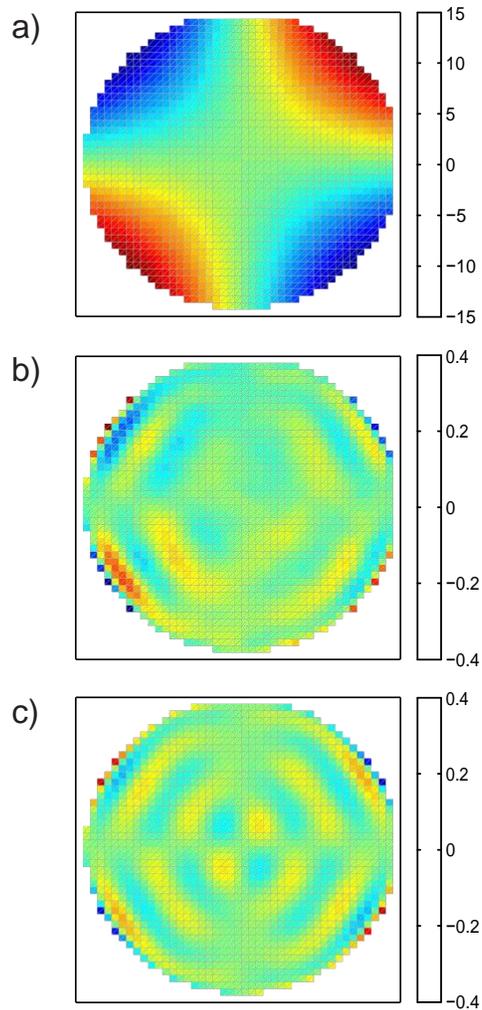

Fig. 5.— a) An astigmatism term having a 30 $\mu$m wavefront amplitude produced by the 91-actuator MLDM. b) The residual wavefront error between a) and the corresponding Zernike term. The rms error is 0.064 $\mu$m. c) The predicted residual wavefront error simulated using a simple model where the influence function of each actuator is aprroximated by a Gaussian function. The predicted residual rms error is 0.059 $\mu$m. Note that this agreement between the predicted and experimental residuals is important since it validates the discussion in section 3. The color-coded amplitudes are in micron units.

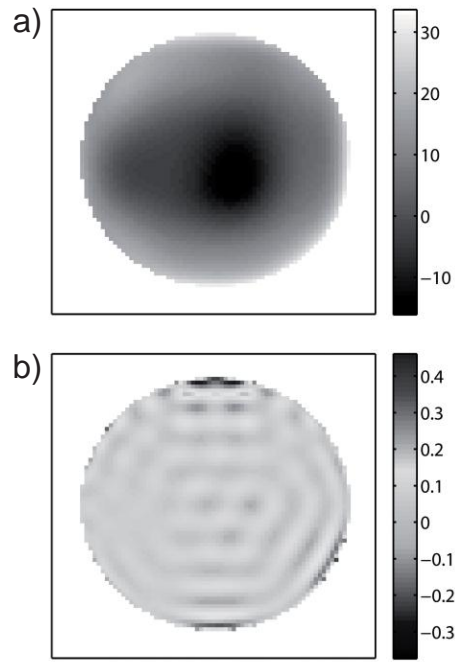

Fig. 6.— a) Reconstructed wavefront for a large astronomical mirror during the early polishing stage (Zernike coefficients by courtesy of the Steward Observatory). b) The predicted residual wavefront error simulated using a 217 actuators MLDM. The color-coded amplitudes are in micron units.

Table 1: Number of actuators requirements for various sizes of the primary mirror (M1) made of facets having 5 and 8-m diameters. The overall diameter of the MLDM in meters is given considering an inter-actuator distance of 5 mm.

| M1 diameter (m) | number of actuators (5-m facets) | MLDM diameter (m) |
|---|---|---|
| 15 | 1519 | 0.25 |
| 25 | 4123 | 0.43 |
| 35 | 8029 | 0.60 |
| M1 diameter (m) | number of actuators (8-m facets) | MLDM diameter (m) |
| 24 | 1519 | 0.25 |
| 40 | 4123 | 0.43 |
| 56 | 8029 | 0.60 |